\documentclass[a4paper]{article}

\usepackage{amsmath, amsfonts}

\usepackage{amssymb}
\usepackage{enumerate}

\providecommand{\keywords}[1]
{
  \small	
  \textbf{\textit{Keywords:}} #1
}
\providecommand{\amsclass}[1]

\newtheorem{theorem}{Theorem}

\newtheorem{lemma}{Lemma}

\title{Energy Escape to Infinity for Systems with Infinite
  Number of Particles}
\author{V.A.~Malyshev$^{1}$,  A.A.~Zamyatin$^{1}$  \\
\small 
\textit{$^{1}$Lomonosov Moscow State University} \\ 
}
\date{}

\begin{document}
\maketitle



\begin{abstract}
Here we consider system of infinite number of particles
where any particle cannot escape to infinity. We define what means
escape of energy to infinity and apply this notion to the case when
constant force  provides energy to one of the particles. 
\end{abstract}

\section{The Model}

Consider the set $X_{\infty}$ of infinite sequences of point particle
coordinates on the real axis 
\begin{equation}
\ldots <x_{-1}<x_{0}<x_{1}<\ldots <x_{k-1}<x_{k}<\ldots ,\label{infinite}
\end{equation}
 We assume that the dynamics is defined by Newton's equations (masses
are assumed to be $1$ )
\begin{equation}
\ddot{x}_{k}=\omega^{2}(x_{k+1}-2x_{k}+x_{k-1})-\omega_{0}^{2}(x_{k}-ka)+f\delta_{k},\label{x_equations}
\end{equation}
where $\delta_{k}=1$ for $k=0$ and $\delta_{k}=0$ for $k\neq0$.
That is the dynamics is defined by the formal potential energy 
\[
U=\frac{\omega^{2}}{2}\sum_{k\in Z}(x_{k+1}-x_{k}-a)^{2}+\frac{\omega_{0}^{2}}{2}\sum_{k\in Z}(x_{k}-ka)^{2}-fx_{0}.
\]

Here we assume that constants $a,\omega,\omega_{0}$ are strictly
positive and external force $f$ is constant. 

It will be convenient to put $q_{k}(t):=x_{k}(t)-ka.$ Then the total
energy at time $t$ can be rewritten as
\[
H(t)=T(t)+U(t),\,\,\,U(t)=W^{int}(t)+W^{0}(t)
\]
where
\[
T(t)=\sum_{k\in Z}\frac{v_{k}^{2}(t)}{2},\quad v_{k}(t)=\frac{dx_{k}}{dt},
\]
\begin{equation}
W^{int}(t)\text{=}\frac{\omega^{2}}{2}\sum_{k\in Z}(q_{k+1}(t)-q_{k}(t))^{2}+\frac{\omega_{0}^{2}}{2}\sum_{k\in Z}q_{k}^{2}(t),\,\,\,W^{0}(t)=-fq_{0}(t).\label{H_t}
\end{equation}
That is the potential energy $U$ is the sum of the interaction energy
$W^{int}$ and the energy of particle $0$ in the external field.
Then the equations (\ref{x_equations}) become
\begin{equation}
\ddot{q}_{k}=\omega^{2}(q_{k+1}-2q_{k}+q_{k-1})-\omega_{0}^{2}q_{k}+f\delta_{k}\label{q_equations}
\end{equation}
Define a matrix $V=\{v_{ij}\}$ with elements
\[
v_{ij}=\begin{cases}
2\omega^{2}+\omega_{0}^{2}, & i=j\\
-\omega^{2}, & |i-j|=1\\
0, & |i-j|>1
\end{cases}
\]
The matrix $V$ defines a positive definite operator $V:l_{2}(Z)\to l_{2}(Z).$ 

We have 
\[
W^{int}(t)\text{=}\frac{1}{2}(Vq(t),q(t))
\]
\[
U(t)=\frac{1}{2}(Vq(t),q(t))-fq_{0}(t)
\]
where $q(t)=\{q_{k}(t),k\in Z\}.$ Equations (\ref{q_equations})
take the form
\[
\ddot{q}=-Vq+fe_{0}
\]
where $e_{0}$ is the vector with coordinates $\{\delta_{k},k\in Z\}.$

We asssume that the initial vector $\phi(0)=\{(q_{k}(0),v_{k}(0)),k\in Z\}\in l_{2}(Z^{2})$.
Then $H(0)$ is finite and it is known (see \cite{DK,D}), that, for
constant $f$, the solution $\phi(t)$ of the system of equations
(\ref{q_equations}) exists in $l_{2}(Z^{2})$ for $0\leq t<\infty$,
and is unique. Then for any $t$ the energy $H(t)=H(0)$.

Define the energy of particle $k$ as 
\[
H_{k}(t)=T_{k}(t)+U_{k}(t),
\]
where
\begin{align*}
  T_{k}(t)&=\frac{v_{k}^{2}(t)}{2},\\
  U_{k}(t) &= \frac{\omega^{2}}{4}((q_{k+1}(t)-q_{k}(t))^{2}+(q_{k}(t)-q_{k-1}(t))^{2})+\frac{\omega_{0}^{2}}{2}q_{k}^{2}(t)-fq_{0}(t)\delta_{k}.
\end{align*}

Denote $H_{A}(t)$ the energy of particles $k$ for all $k\in A$,
that is
\begin{align*}
  H_{A}(t)&=T_{A}(t)+U_{A}(t),\\
  U_{A}(t)&=\sum_{k\in A}U_{k}(t),T_{A}(t)=\sum_{k\in A}T_{k}(t).
\end{align*}

We shall say that all energy escapes to infinity if for any $N>0$
\begin{equation}
\lim_{t\to\infty}H_{[-N,N]}(t)=0. \label{en_1}
\end{equation}
We shall say that no energy escapes to infinity if for any $N>0$
there is a bound uniform in $t$ 
\begin{equation}
H_{(-\infty,-N)}(t)+H_{(N,\infty)}(t)<C_{N} \label{en_2}
\end{equation}
and such that $C_{N}\to0$ when $N\to\infty$. 

If for any $k$ 
\[
H_{k}(t)\to h_{k}
\]
for some fixed $h_{k}$ then only the part 
\[
H(0)-\sum_{k\in Z}h_{k}
\]
of energy escapes to infinity. 

\section{Constant external force}

\paragraph{Results}

Here we always consider initial conditions in $l_{2}(Z)$.

\begin{lemma}\label{l0}
Let $f=0$. Then, for any initial conditions we have: for any $k$
\[
\lim_{t\to\infty}q_{k}(t)=0,\quad \lim_{t\to\infty}\dot{q}_{k}(t)=0
\]
and all energy escapes to infinity in the sense of formula (\ref{en_1}).
\end{lemma}

If $f\neq0$, consider particular solution $\{q_{k}(t)=\xi_{k}\}\in l_{2}(Z),v_{k}(t)=0$
of nonhomogeneous system, which does not depend on time. Here $\xi_{k}$
are constants satisfying the following system of equations
\begin{equation}
\omega^{2}(\xi_{k+1}-2\xi_{k}+\xi_{k-1})-\omega_{0}^{2}\xi_{k}+f\delta_{k}=0. \label{ksi_equations}
\end{equation}
In the matrix form we have
\[
V\xi=fe_{0}
\]
where vector $\xi=\{\xi_{k},k\in Z\}.$

\begin{lemma} \label{l1}
System (\ref{ksi_equations}) has unique solution $\xi=fV^{-1}e_{0}\in l_{2}(Z$
with coordinates
\begin{equation}
\xi_{k}=\frac{1}{\pi}\int_{-\pi/2}^{\pi/2}\frac{f\cos(2k\varphi)}{4\omega^{2}\sin^{2}\varphi+\omega_{0}^{2}}d\varphi . \label{ksi_k}
\end{equation}
Moreover, for any $p>0$ there exists constant $C(p)$ such that for
any $k$
\[
|\xi_{k}|\leq C(p)k^{-p}.
\]
\end{lemma}

Denote
\[
U_{\xi}=\sum_{k\in Z}U_{\xi,k},\,\,\,U_{\xi,k}=\frac{\omega^{2}}{4}(\xi_{k+1}-\xi_{k})^{2}+\frac{\omega_{0}^{2}}{2}\xi_{k}^{2}-f\xi_{0}\delta_{k}
\]
the potential energy of this solution. We have
\[
U_{\xi}=\frac{1}{2}(V\xi,\xi)-f\xi_{0}=\frac{1}{2}(fe_{0},\xi)-f\xi_{0}=-\frac{1}{2}f\xi_{0}=-Cf^{2}<0
\]
where
\[
C=\frac{1}{2\omega_{0}\sqrt{4\omega^{2}+\omega_{0}^{2}}}>0.
\]

Equations (\ref{q_equations}) with initial conditions $q_{k}(0),v_{k}(0)$  have unique solution $q_{k}(t)$ $=\xi_{k}+\zeta_{k}(t)$, where
the vector $\zeta(t)=\{\zeta_{k}(t)\}$ is the solution of homogeneous
equation 
\[
\ddot{\zeta}=-V\zeta
\]
 with initial conditions 
\begin{equation}
\zeta_{k}(0)=q_{k}(0)-\xi_{k},v_{k}(0). \label{initial}
\end{equation}
Note that for solution $\zeta_{k}(t)$ the potential energy is equal
to
\[
W^{int}(t)=\frac{1}{2}(V\zeta(t),\zeta(t))=\frac{1}{2}(V(q(t)-\xi),q(t)-\xi)=
\]
\[
=\frac{1}{2}(Vq(t),q(t))-\frac{1}{2}(Vq(t),\xi)-\frac{1}{2}(q(t),V\xi)+\frac{1}{2}(V\xi,V\xi)=
\]
\[
=\frac{1}{2}(Vq(t),q(t))-(q(t),V\xi)+\frac{1}{2}(V\xi,V\xi)=
\]
\[
= \frac{1}{2}(Vq(t),q(t))-fq_{0}(t)+\frac{1}{2}f\xi_{0}=U(t)+Cf^{2}.
\]
In particular, it follows, that
\[
U_{\xi}+Cf^{2}=0.
\]
The total energy $H^{hom}(t)$ of homogeneous system is
\begin{equation}
H^{hom}(t)=W^{int}(t)+T(t)=U(t)+T(t)+Cf^{2}=H(t)+Cf^{2}\label{hom}
\end{equation}
as $H(t)=U(t)+T(t).$

\begin{theorem}\label{th1} 
Let $f\neq0$. Then for all $k$ we have 
\[
\lim_{t\to\infty}q_{k}(t)=\xi_{k},\:\lim_{t\to\infty}\dot{q}_{k}(t)=0
\]
and the amount of energy that escapes to infinity is
\[
\lim_{N\to\infty}\lim_{t\to\infty}(H_{(-\infty,-N)}(t)+H_{(N,\infty)}(t))=H(0)+Cf^{2}=H^{hom}(0).
\]
So the amount of energy that escapes to infinity is equal to the initial
energy of the homogeneous system with initial conditions (\ref{initial}).
\end{theorem} 

In particular, for zero initial conditions $q_{k}(0)=0,v_{k}=0$ this
energy is $Cf^{2}$, that is equal to minus potential energy of the
$\xi$ configuration. For initial conditions $q_{k}(0)=\xi_{k},v_{k}=0$,
no energy escapes to infinity.

\paragraph{Proofs}

The proof of theorem \ref{th1} is based on lemmas \ref{l0} and \ref{l1}. 

\paragraph{Proof of lemma \ref{l1}.}

To get formula (\ref{ksi_k}) we define Fourier transform 
\[
X(\varphi)=\sum_{k\in Z}\xi_{k}e^{ik\varphi}
\]
and come to Fourier transform in equations (\ref{ksi_equations}).
Then we find
\[
X(\varphi)=\sum_{k\in Z}\xi_{k}e^{ik\varphi}=\frac{f}{\Omega^{2}(\varphi)}
\]
where
\[
\Omega^{2}(\varphi)=2\omega^{2}(1-\cos\varphi)+\omega_{0}^{2}=4\omega^{2}\sin^{2}\frac{\varphi}{2}+\omega_{0}^{2}.
\]
So 
\[
\xi_{k}=\frac{1}{2\pi}\int_{-\pi}^{\pi}\frac{fe^{-ik\varphi}}{\Omega^{2}(\varphi)}d\varphi=\frac{1}{2\pi}\int_{-\pi}^{\pi}\frac{fe^{-ik\varphi}}{4\omega^{2}\sin^{2}\frac{\varphi}{2}+\omega_{0}^{2}}d\varphi=
\]
\[
=\frac{1}{\pi}\int_{-\pi/2}^{\pi/2}\frac{fe^{-i2k\varphi}}{4\omega^{2}\sin^{2}\varphi+\omega_{0}^{2}}d\varphi.
\]
One can write also
\[
\xi_{k}=\frac{1}{\pi}\int_{-\pi/2}^{\pi/2}\frac{f\cos(2k\varphi)}{4\omega^{2}\sin^{2}\varphi+\omega_{0}^{2}}d\varphi.
\]

Integrating by parts the required number of times and using the periodicity
of the integrand one can get $\xi_{k}=O(k^{-p}),p>0.$ Lemma is proved.

\paragraph{Proof of lemma \ref{l0}.}

Define Fourier transforms
\[
Q(t,\varphi)=\sum_{k\in Z}q_{k}(t)e^{ik\varphi},\:\dot{Q}(t,\varphi)=\sum_{k\in Z}\dot{q}_{k}(t)e^{ik\varphi},\:\varphi\in[-\pi,\pi].
\]
We denote by $\dot{Q},\ddot{Q}$ time derivatives. Coming to Fourier
transform in (\ref{q_equations}) with $f=0$ we get the following
differential equation
\begin{equation}
\ddot{Q}(t,\varphi)=-\Omega^{2}(\varphi)Q(t,\varphi)\label{d_e}
\end{equation}
with initial conditions
\[
Q(0,\varphi)=\sum_{k\in Z}q_{k}(0)e^{ik\varphi}, \quad \dot{Q}(0,\varphi)=\sum_{k\in Z}\dot{q}_{k}(0)e^{ik\varphi}.
\]
 The solution of equation (\ref{d_e}) is
\begin{align}
Q(t,\varphi) & =Q(0,\varphi)\cos(\Omega t)+\dot{Q}(0,\varphi)\frac{\sin(\Omega t)}{\Omega},\label{f_tr}\\
\dot{Q}(t,\varphi) & =-Q(0,\varphi)\Omega\sin(\Omega t)+\dot{Q}(0,\varphi)\cos(\Omega t),\nonumber 
\end{align}
where 
\[
\Omega=\Omega(\varphi)=\sqrt{4\omega^{2}\sin^{2}\frac{\varphi}{2}+\omega_{0}^{2}}.
\]
Inverting Fourier transform (\ref{f_tr}) we get 
\[
q_{k}(t)=\frac{1}{2\pi}\int_{-\pi}^{\pi}(Q(0,\varphi)\cos(\Omega t)+\dot{Q}(0,\varphi)\frac{\sin(\Omega t)}{\Omega})e^{-ik\varphi}d\varphi,
\]
or after the change of variable we have
\[
\zeta_{k}(t)=\frac{1}{\pi}\int_{-\pi/2}^{\pi/2}(Q(0,2\varphi)\cos(\Omega(2\varphi)t)+\dot{Q}(0,2\varphi)\frac{\sin(\Omega(2\varphi)t)}{\Omega(2\varphi)})e^{-i2k\varphi}d\varphi.
\]
To finish the proof we need the following result. 

\begin{lemma}\label{l3}
The following asymptotic formula holds 
\[
\zeta_{k}(t)=\frac{1}{\sqrt{t}}(C_{1}\cos(t\omega_{0}+\frac{\pi}{4})+(-1)^{k}C_{2}\cos(t\sqrt{4\omega^{2}+\omega_{0}^{2}}-\frac{\pi}{4}))+
\]
\[
+\frac{1}{\sqrt{t}}(S_{1}\sin(t\omega_{0}+\frac{\pi}{4})+(-1)^{k}S_{2}\sin(t\sqrt{4\omega^{2}+\omega_{0}^{2}}-\frac{\pi}{4}))+O(t^{-\frac{3}{2}})
\]
where
\[
C_{1}=\sqrt{\frac{\omega_{0}}{2\pi\omega^{2}}}Q(0,0), \quad C_{2}=\sqrt{\frac{\sqrt{4\omega^{2}+\omega_{0}^{2}}}{2\pi\omega^{2}}}Q(0,\pi)
\]
\[
S_{1}=\sqrt{\frac{1}{2\pi\omega\omega_{0}}}\dot{Q}(0,0), \quad S_{2}=\sqrt{\frac{1}{2\pi\omega\sqrt{4\omega^{2}+\omega_{0}^{2}}}}\dot{Q}(0,\pi)
\]
as $t\to\infty.$
\end{lemma}

{ \it Proof of lemma \ref{l3}.} It was proved in \cite{LM}. We will give the proof for completeness. 

  We will use the stationary phase method. (See Fedoruk, p.\thinspace 102). In
our case phase function is
\[
S(\varphi)=\Omega(2\varphi)=\sqrt{4\omega^{2}\sin^{2}\varphi+\omega_{0}^{2}},\varphi\in[-\frac{\pi}{2},\frac{\pi}{2}]
\]
where
\[
\omega_{0}\leq S(\varphi)\leq\sqrt{4\omega^{2}+\omega_{0}^{2}}.
\]
We have stationary points $0,\pm\frac{\pi}{2}$ and 
\begin{align*}
  S_{\varphi\varphi}(0)&=\frac{4\omega^{2}}{\omega_{0}},\\
  S_{\varphi\varphi}(\frac{\pi}{2})&=-\frac{4\omega^{2}}{\sqrt{4\omega^{2}+\omega_{0}^{2}}}.
\end{align*}
For
\[
a_{k}(t)=\frac{1}{\pi}\int_{-\pi/2}^{\pi/2}Q(0,2\varphi)\cos(S(\varphi)t)e^{-i2k\varphi}d\varphi
\]
due to periodicity of the integrand (with period $\pi$) we get
\[
a_{k}(t)=\frac{1}{\pi}\int_{-\pi/2+\epsilon}^{\pi/2+\epsilon}Q(0,2\varphi)\cos(tS(\varphi))e^{-i2k\varphi}d\varphi.
\]
So we have two stationary points $0,\frac{\pi}{2}$ inside the segment
$[-\frac{\pi}{2}+\epsilon,\frac{\pi}{2}+\epsilon]$. Further on, we
write
\[
a_{k}(t)=\frac{1}{2\pi}\int_{-\pi/2+\epsilon}^{\pi/2+\epsilon}Q(0,2\varphi)(e^{itS(\varphi)}+e^{-itS(\varphi)})e^{-i2k\varphi}d\varphi,
\]
and apply formula from Fedoruk (p. 102). This gives
\[
a_{k}(t)=\frac{1}{\sqrt{t}}(C_{1}\cos(t\omega_{0}+\frac{\pi}{4})+(-1)^{k}C_{2}\cos(t\sqrt{4\omega^{2}+\omega_{0}^{2}}-\frac{\pi}{4}))+O(t^{-\frac{3}{2}}).
\]
Similar for 
\[
b_{k}(t)=\frac{1}{\pi}\int_{-\pi/2+\epsilon}^{\pi/2+\epsilon}\dot{Q}(0,2\varphi)\frac{\sin(S(\varphi)t)}{S(\varphi)}e^{-i2k\varphi}d\varphi
\]
we have asymptotic formula
\[
b_{k}(t)=\frac{1}{\sqrt{t}}(S_{1}\sin(t\omega_{0}+\frac{\pi}{4})+(-1)^{k}S_{2}\sin(t\sqrt{4\omega^{2}+\omega_{0}^{2}}-\frac{\pi}{4}))+O(t^{-\frac{3}{2}}).
\]
The lemma is proved.

So lemma \ref{l0} follows from lemma \ref{l3}.

To finish the proof of the theorem note that by lemma \ref{l0}
\[
\lim_{t\to\infty}H_{[-N,N]}(t)=U_{\xi,[-N,N]}=\sum_{k\in[-N,N]}U_{\xi,k}.
\]
As $H(t)=H(0),$ then
\[
\lim_{t\to\infty}(H_{(-\infty,-N)}(t)+H_{(N,\infty)}(t))=H(0)-U_{\xi,[-N,N]}=
\]
\[
H(0)-U_{\xi}+U_{\xi,(-\infty,-N)}+U_{\xi,(N,\infty)}=H(0)+Cf^{2}+U_{\xi,(-\infty,-N)}+U_{\xi,(N,\infty)}
\]
and
\[
\lim_{N\to\infty}(U_{\xi,(-\infty,-N)}+U_{\xi,(N,\infty)})=0.
\]
By (\ref{hom}),
\[
H(0)+Cf^{2}=H^{hom}(0).
\]


\begin{thebibliography}{5}
  
\bibitem{LM}
A.~Lykov and M.~Melikyan  (2019)
\newblock Long time behavoir of infinite
harmonic chain with $l_{2}$ initial conditions. 
\newblock {\it Structure of mathematical
physics} {\bf 2} (1), 5--26 .

\bibitem{DK}
Ju.L.~Daleckii and M.G.~Krein (1974) 
\newblock {\it Stability of Solutions
of Differential Equations in Banach Space}. AMS. 

\bibitem{D} 
K.~Deimling (1977) 
\newblock {\it Ordinary Differential Equations in
Banach Spaces}. 
Springer.
\end{thebibliography}
\end{document}